
\magnification=\magstep 1
\topskip 2 true cm
\font\bigtenrm=cmbx10 scaled\magstep 3
\parskip =2pt
{\rightline{FISIST/12-95/CFIF}}
{\rightline{hep-ph/9507299}}
\bigskip\bigskip\bigskip\bigskip
{\centerline{\bigtenrm Predicting  V${\bf_{CKM}}$
with}}\bigskip
{\centerline{\bigtenrm Universal Strength of Yukawa
Couplings}}
\bigskip\bigskip
\bigskip\bigskip \bigskip\bigskip
{\centerline{{\bf G. C. Branco} and {\bf J. I. Silva-Marcos}}}
\bigskip
{\centerline{\it Departamento de F¡sica and CFIF}}
{\centerline{\it Instituto Superior T'cnico}}
{\centerline{\it Avenida Rovisco Pais, 1096 Lisboa Codex,
Portugal}}
\bigskip\bigskip\bigskip\bigskip
{\centerline{\bf ABSTRACT}}
\bigskip
We study the question of calculability of the
Cabibbo-Kobayashi-Maskawa (CKM) matrix elements
within the framework of universal strength for
Yukawa couplings (USY). We first classify all solutions
leading to $m_u=m_d=0$ within USY and then
suggest a highly predictive ansatz
where all the moduli of the CKM matrix elements are
correctly predicted in terms of quark mass ratios,
{\it with no free parameters}.
\vfill\eject

{\bf 1. Introduction}. Understanding
the  structure of fermion masses and mixings is one of
the outstanding problems in particle physics.
In the standard model (SM) both quark
masses and mixings are arbitrary, since gauge invariance
does not constrain the flavour structure of Yukawa
couplings.

Sometime ago, it was suggested [1] that the Yukawa
couplings of the SM have exact universal strength
leading to quark mass matrices of the form:
$$
M_u\,=\, c_u\,\left[e^{i\phi^u_{ij}}\right]\qquad ;
\qquad
M_d\,=\, c_d\,
\left[e^{i\phi^d_{ij}}\right]\eqno(\, 1.1\,)
$$
where $c_u$ and $c_d$ are real numbers. Within the
framework of the standard Higgs mechanism and given
the quark mass spectrum, the hypothesis of universal
strength for Yukawa couplings (USY) requires a minimum of
two Higgs doublets, with the up and down quarks acquiring
their masses through the couplings of the two different
doublets, $\Phi_u$ and $\Phi_d$. The constants $c_u$ and
$c_d$ are given by $c_u\,=\,|g_Y\,\, v_u|$ and
$c_d\,=\,|g_Y\,\, v_d|$, where $g_Y$ is the universal
strength of the Yukawa couplings and $v_u$, $v_d$ stand
for $<\Phi_u>$, $<\Phi_d>$ respectively. It is worth
noting
that in the SM, with one Higgs doublet, Yukawa couplings
are the only ones which can be complex; all other
couplings
have to be real as a result of Hermiticity. The same
applies to the SM with two Higgs doublets, provided the
selective Yukawa couplings of $\Phi_u$ and $\Phi_d$
leading to natural flavour conservation, result from a
symmetry of the Lagrangean [2]. The USY hypothesis has the
appeal of suggesting that the observed rich spectrum of
quark masses and mixings simply results from the fact that
Yukawa couplings can be complex, with universal strength,
but undetermined phase.

It has been shown [1,3] that within the USY hypothesis,
one can fit all the experimental values of
quark masses and mixings. We find it remarkable
that this is possible, keeping exact
universality of strength of Yukawa couplings.
However, a drawback of the USY hypothesis is
the fact that it contains a large number
of free parameters which weakens its predictive power.

In this paper, we will address the question
of whether it is possible to achieve
predictability of the CKM matrix
elements within the USY framework.
First we study the limit
where the first generation of quarks is massless and show
that solutions leading to vanishing $m_u$ and $m_d$ can
be classified into two classes. We then find an exact
analytical solution for $V_{CKM}$ in the limit
$m_u\,=\, m_d\,=\, 0$ and show that the main features of
$V_{CKM}$ are correctly predicted. Inspired by this
analysis, we propose a specific ansatz, within the USY
hypothesis, which predicts $V_{CKM}$ in terms of quark
mass ratios with no free parameters. The moduli of all
CKM matrix elements are correctly predicted. However,
within this specific ansatz, the implied strength of
CP violation through the KM mechanism is not sufficient
to account for the observed CP violation in the Kaon
sector,
thus suggesting significant contributions to $\epsilon$
from physics beyond the SM.

{\bf 2. Characterisation of parameter space}. Next, we
will
characterize the parameter space, first eliminating the
unphysical phases and then separating the remaining ones
into those which do not enter in the determination of the
quark mass eigenvalue spectrum, but affect $V_{CKM}$ and
those which enter both in the determination of the quark
mass spectrum and the evaluation of $V_{CKM}$.

By making phase transformations on the right-handed
quark fields $d_R^i$ and $u_R^i$, the mass matrices
$M_u$
and $M_d$ can, without loss of generality, be written in
the
form:
$$
M=c \quad K^{\dagger}\cdot\left(\matrix{e^{ip} &e^{ir}
&1\cr
e^{iq} &1 &e^{it}\cr
1&1&1\cr}\right)\cdot K  \eqno(\, 2.1\, )
$$
where $K =$diag$(1,e^{i\alpha_{1}},e^{i\alpha_{2}})$.
We have omitted the subscripts $u$ and $d$
throughout, since both $M_u$ and $M_d$ can be put in the
above form. The charged currents remain diagonal and
real, so Eq.(2.1) just reflects a choice of weak basis. The
advantage of writing $M_u$ and $M_d$ in this basis is
that
the phases $\alpha_i$ entering in the diagonal
matrices,
$K_{u,d}$ do not affect the quark mass spectrum, which
only depends on the phases $\{p,q,r,t\}$. However, the
phases $\alpha_i$ do enter in the evaluation of the CKM
matrix. It is useful to introduce the Hermitian matrices:
$$
{\tilde H}_u\,=\, {1\over 3c_u^2}\quad K_u^\dagger\,\,
H_u\,\, K_u\qquad ;\qquad
{\tilde H}_d\,=\, {1\over 3c_d^2}\quad K_d^\dagger\,\,
H_d\,\, K_d\eqno(\, 2.2\, )
$$
where $H\,=\, M\,\, M^\dagger$. The matrices
${\tilde H}_u$ and ${\tilde H}_d$ can be written in the
form:

$$
{\tilde H}\,=
\,\left(\matrix {1&{e^{i(p-q)}+e^{ir}+e^{-it}\over 3}&
{e^{ip}+e^{ir}+1\over 3}\cr &&\cr
{e^{-i(p-q)}+e^{-ir}+e^{it}\over 3} &1&
{e^{iq}+e^{it}+1\over 3}\cr &&\cr
{e^{-ip}+e^{-ir}+1\over 3}&
{e^{-iq}+e^{-it}+1\over 3}&1\cr}\right)\eqno(\, 2.3\, )
$$
The eigenvalues of ${\tilde H}$ are related to the square
quark masses by $\lambda_{i}=3m_i^2 /
[ m_{3}^2+m_{2}^2+m_1^2]$.
The coefficients of the characteristic equation for
${\tilde H}$ can be expressed in terms of $p$, $q$, $r$
and $t$:

$$
\eqalign {&tr({\tilde H})=
\lambda_{1}+\lambda_{2}+\lambda_{3}=3\cr
&{\cal X}({\tilde H})=
\lambda_{1}\lambda_{2}+\lambda_{1}\lambda_{3}+
\lambda_{2}\lambda_{3}=
\,\,{4\over 9}\biggl[\sin^2({{p\over 2}})+
\sin^2({{q\over 2}})+\sin^2({{r\over 2}})
+\sin^2({{t\over 2}})+\cr
&\qquad\sin^2({r+t\over 2})+
\sin^2({{p-r\over 2}})+\sin^2({{q-t\over 2}})
+\sin^2({{p-q-r\over 2}})+\sin^2({p-q+t\over 2})\biggr]\cr
&det({\tilde H}) =
\lambda_{1}\lambda_{2}\lambda_{3}=
{1\over 3}{\cal X}({\tilde H}) -{4\over 27}
\biggl[ \sin^2({{p-q\over 2}})+\sin^2({{p+t\over 2}})
+\sin^2({{q+r\over 2}})+\cr
&\qquad \sin^2({{p-r-t\over 2}})
+\sin^2({{q-r-t\over 2}})
+\sin^2({p-q-r+t\over 2})\biggr ]\cr}  \eqno(\, 2.4\, )
$$

Since the observed mass hierarchy leads to the
constraint ${\cal X}({\tilde H})<<\, 1$ and given
the fact that ${\cal X}({\tilde H})$ in Eq.(2.4) is
the sum of positive definite
quantities, each one of the parameters $p^2$, $q^2$,
$r^2$
and $t^2$ has to be small compared to the unity. In
order
to get an idea of the size of this bound, let us consider
the physically interesting limit $r_{ _u}=t_{ _u}=r_{
_d}=t_{ _d}=0$,
where the masses of the first family vanish. In this
limit
one has:

$$
\sin^2({p_{ _d}\over 2})+\sin^2({q_{ _d}\over 2})+
\sin^2({p_{ _d}-q_{ _d}\over 2})=\,\,
{9\over 8}\,{\cal X}_d({\tilde H_d})=
\,\,{81\over
8}\,\,{m_s^2m_b^2\over\left[m_s^2+m_b^2\right]^2}
\eqno(\, 2.5\, )
$$

\noindent
which constraints both $|p_{ _d}|$ and $|q_{ _d}|$ to be
less
than $(9/2)(m_s/m_b)$. Obviously, analogous constraints
hold
for the up quark sector.

We have seen that as a result of the quark hierarchy the
phases
$p$, $q$, $r$ and $t$ have to be all small. We will show
next
that the general pattern of the CKM matrix constrains the
remaining phases to be small. This can be seen by the
following argument. In general, it can be shown that in
the
limit where all quark masses vanish, except $m_t$ and
$m_b$,
the following relation holds:

$$
{tr(H_d)tr(H_u)-tr(H_dH_u)\over tr(H_d)tr(H_u)}\,\,
=\,\,1-|V_{tb}|^2\eqno(\, 2.6\, )
$$

\noindent
This relation is obtained by noting that all terms
entering in the left hand side of Eq.(2.6) and (2.8)
are weak-basis
invariants. Therefore, one can choose to evaluate them
in the basis where $H_u\,=\, {\rm
diag.}(0,0,m_t^2)$.
In this basis,
$H_d\,=\, V_{CKM}\cdot{\rm diag.}(0,0,m_b^2)\cdot
V_{CKM}^\dagger$. Using
these expressions for $H_u$ and $H_d$, the result of
Eq.(2.6) follows. In the USY framework, the limit
$m_u=m_c=m_d=m_s=0$, corresponds, to having $p=q=r=t=0$
in
both the up and down quark sectors. The matrices $M_u$
and
$M_d$ have then the form:

$$
M_u\,\,=\,\, c_u\quad K_u^\dagger\,\,\Delta\,\, K_u
\qquad ; \qquad
M_d\,\,=\,\, c_d\quad K_d^\dagger\,\,\Delta\,\,
K_d\eqno(\, 2.7\, )
$$

\noindent
where $\Delta$ is the democratic matrix [4], with
all entries equal to the unity. The Hermitian
matrices $H_u$ and $H_d$ are given by:

$$
H_u\,\,=\,\, 3c_u^2\quad K_u^\dagger
\,\,\Delta\,\, K_u\qquad ;\qquad
H_d\,\,=\,\, 3c_d^2\quad K_d^\dagger
\,\,\Delta\,\, K_d\eqno(\, 2.8\, )
$$

\noindent
where we have taken into account that
$\Delta^2=\,\, 3\Delta$. Using Eq.(2.6) one obtains:

$$
\sin^2({\psi_1\over 2})+\sin^2({\psi_2\over 2})+
\sin^2({\psi_1-\psi_2\over 2})=\,\,
{9\over 4}\,\,\left[1-|V_{tb}|^2\right]\eqno(\, 2.9\, )
$$

\noindent
where $\psi_i\equiv\alpha_i^u-\alpha_i^d$. The fact that
experimentally $\left(1-|V_{tb}|^2\right)<<1$, constrains
the phases $\psi_i$ to be small. At this point, the
following comment is in order. The result of Eq.(2.6) is
valid for any arbitrary matrices $H_u$ and $H_d$ when
they approach the rank one limit, with all masses except
$m_t$ and $m_b$ vanishing. The democratic matrices, with
$H_u$ and $H_d$ proportional to $\Delta$, are a special
case of rank one matrices, where the phases $\psi_i$
vanish.

{\bf 3. USY in the limit ${\bf m_u=m_d=0}$}.
Experimentally, it
is known that the first generation of quarks has much
smaller masses than the other two, which provides
motivation to study the above limit. We will show that
in
the
USY framework, all solutions with $ m_u=m_d=0$ can be
classified into two classes, which correspond to simple
choices for the parameters $\{p,q,r,t\}$.

The equation
$det(M)=0$, with $M$ given by Eq.(2.1), leads to the
following
relations:

$$\cases {\cos(p)-\cos(p+t)+\cos(q)-\cos(q+r)=\,\,
1-\cos(r+t)\cr
\sin(p)-\sin(p+t)+\sin(q)-\sin(q+r)=\,\,-\sin(r+t)\cr}
\eqno(\, 3.1\, )
$$

\noindent
Using trigonometric identities, such as
$\cos(A)-\cos(B)=-2\sin({A-B\over 2})\sin({A+B\over 2})$,
one can write
Eq(3.1) as:

$$
\cases {\sin({t\over 2})\,\,\sin({p}+{t\over 2})+
\sin({r\over 2})\,\,\sin(q+{r\over 2})=
\sin({r+t\over 2})\,\,\sin({r+t\over 2})\cr \cr
\sin({t\over 2})\,\,\cos({p}+{t\over
2})+\sin({r\over2})\,\,
\cos({q}+{r\over 2})=\sin({r+t\over2})\,\,
\cos({r+t\over 2})\cr}\eqno(\, 3.2\, )
$$

\noindent
Squaring and adding the two Eqs.(3.2), one finally obtains:

$$
\sin({r\over 2}).\sin({t\over 2}).\sin({p-q-r\over 2})
.\sin({p-q+t\over 2})=\, 0
\eqno(\, 3.3\, )
$$

\noindent
 From Eq.(3.1) and Eq.(3.3), one readily concludes that only
the following solutions exist;
we divide them into two classes:

$$
\eqalign{
&{\rm Class\,\, I}\quad\cases{a)\cr b)\cr c)\cr}\matrix{
& p=0\, ,&t=q&;\,\, q,r\,\,{\rm free}\cr
& r=0\, ,&t=0 &;\,\, p,q\,\,{\rm free}\cr
& r=p\, ,&q=0 &;\,\, p,t\,\,{\rm free}\cr}
\cr
&{\rm Class\,\, II}\quad\cases{a)\cr b)\cr c)\cr}\matrix{
& q=0\, ,&t=0 &;\,\, p,r\,\,{\rm free}\cr
& p=0\, ,&r=0 &;\,\, q,t\,\,{\rm free}\cr
& p=q+r\, ,&t=-r &;\,\, p,r\,\,{\rm free}\cr} \cr}
\eqno(\, 3.4\, )
$$

\noindent
It is trivial to check that for these simple
cases, $m_u=m_d=0$. The interest of the above analysis,
is that it shows that Eqs.(3.4) are
${\underline {\rm all}}$ the ${\underline{\rm solutions}}$
of the equation $det(M)=0$, in the USY framework.
The reason why it is possible to classify all solutions
into ${\underline {\rm two}}$ classes, has to do with
the fact that two solutions within the same class can be
transformed into each other by pure phase unitary
matrices,
combined with permutations. As an example, let us
consider,
e.g., the solutions (a) and (b) of Class I:

$$
M^I_{(a)}\,=\,\left(\matrix{1&e^{ir}&1\cr & & \cr
e^{iq}&1&e^{iq}\cr & & \cr
1&1&1\cr}\right)\qquad ;\qquad
M^I_{(b)}\,=\,\left(\matrix{e^{ip} &1 &1\cr  & & \cr
e^{iq} &1 &1\cr  & & \cr
1 &1&1\cr}\right)\eqno(\, 3.5\, )
$$
By applying to $M^I_{(b)}$ the transformation,

$$
M^I_{(b)}\quad\longrightarrow\quad
K\,\, P_{23}\,\,\,\,\,\, M^I_{(b)}\,\,
\,\,\,\, P_{12}\,\, K^\dagger \,\, ,
\eqno(\, 3.6\, )
$$

\noindent where $K={\rm diag.}(1,e^{iq},1)$ and $P_{12}$
and $ P_{23}$ are permutations of the family indices,
i.e.,
$$
P_{12}= \,\left(\matrix{0 &1 &0\cr
1 &0 &0\cr
0 &0&1\cr}\right)\qquad ;\qquad
P_{23}= \,\left(\matrix{1 &0 &0\cr
0 &0 &1\cr
0 &1&0\cr}\right)\,\, ,\eqno(\, 3.7\, )
$$

\noindent one obtains a matrix of the form $M^I_{(a)}$.
The transformation of Eq.(3.6) is of course
done both in the up and
down quark sectors. It is easy to see that $M^I_{(a)}$
and
$M^I_{(b)}$ are physically equivalent and lead to the
same
CKM matrix.

We will show that
the two classes of solutions have different
physical implications. For simplicity, let us consider
that $K_u=K_d=1$. It can then be verified that, in
the solutions of Class II, the first generation of
quarks decouples from the other two, so that only the
second and third generation are mixed in the CKM matrix.
This is the situation previously encountered in the
literature [5]. On the contrary Class I solutions have the
novel feature that the full CKM matrix is generated,
even in the limit $m_u=m_d=0$.

Next, we will analyse in detail the solutions of Class I.
We will
derive an exact analytical expression
for $V_{CKM}$ and show how
some of its main features can be understood.

{{\bf Evaluation of the CKM matrix}}.
For definiteness, we will consider both for the
up and down quark sectors a solution I(a), corresponding
to mass matrices of the form:
$$
M_{u,d}\,=\,\,\,
c_{u,d} \quad\left(\matrix{1&e^{ir}&1\cr & & \cr
e^{iq}&1&e^{iq}\cr & & \cr
1&1&1\cr}\right)_{u,d}\eqno(\, 3.8\, )
$$
It is convenient to make a change of weak-basis under
which:
$$
\eqalign{&
H_u\quad\longrightarrow\quad H'_u\,=\, F^\dagger\,\,
H_u\,\, F\cr &\cr &
H_d\quad\longrightarrow\quad H'_d\,=\, F^\dagger\,\,
H_d\,\, F\cr}\eqno(\, 3.9\, )
$$
\noindent where F is given by
$$
F=\left(\matrix{1 \over{\sqrt 2}&-1 \over{ \sqrt 6}
&1\over{ \sqrt 3}\cr
0 &2\over{\sqrt 6} &1\over{\sqrt 3}\cr
-1\over{\sqrt 2} &-1\over{\sqrt 6} &1\over {\sqrt 3}\cr}
\right)\eqno(\, 3.10\, )
$$
The transformation of Eq.(3.9) corresponds to changing from
the "democratic basis" to the "heavy basis".
One can then find an
exact analytical solution for the matrices $U_u$ and
$U_d$
which diagonalize $H'_u$ and $H'_d$:

$$
U^\dagger_d\,\,\,\, H'_d\,\,\,\, U_d\,\,=\,\,{\rm
diag.}
(m_d^2,m_s^2,m_b^2)\eqno(\, 3.11\, )
$$

\noindent with analogous expressions for $H'_u$. The CKM
matrix is then given by $V_{CKM}= U_u^\dagger\,\, U_d$.
The analytical solutions for $U_u$ and $U_d$ can be
written as:

$$
U\,\,=\,\, V_K\,\, V_\theta\,\, V_\phi\eqno(\, 3.12\, )
$$

\noindent where $V_K\,=\, F^\dagger\,\, K\,\, F$, with
$K={\rm diag.}(1,e^{iq},1)$. The matrices $V_\theta$ and
$V_\psi$ are unitary transformations in the $(1,2)$ and
$(2,3)$ generation space, given by:
$$
V_\theta\,\,=\,\, \left(\matrix{a^*/n&\epsilon/n&0\cr
& & \cr -\epsilon^*/n&a/n&0 \cr& & \cr 0&0&1\cr}\right)
\qquad ;\qquad V_\psi\,\,=\,\, \left(\matrix{1&0&0\cr
& & \cr 0&\cos(\psi)&\sin(\psi)\,\, e^{-i\gamma} \cr
&&\cr
0&-\sin(\psi)\,\,
e^{i\gamma}&\cos(\psi)\cr}\right)\eqno(\, 3.13\, )
$$
where, $\epsilon$, $a$, $n$, $\psi$ and $\gamma$ are
simple functions of $q$ and $r$:
$$
\epsilon\,=\, {e^{ir}-1\over\sqrt 2}\,\, ;\,\,
a\,=\, {2e^{-iq}-e^{ir}-1\over\sqrt 6}\,\, ;\,\,
n^2\,\,=\,\, |\epsilon|^2+|a|^2 \,\, ;\,\,
\tan(2\psi)\,=\, {2\sqrt{3} \, |b|\, n\over
6+3|b|^2-n^2}\eqno(\, 3.14\, )
$$
with $b\,=\,(1/3) (1+e^{ir}+e^{-iq})$ and
$\gamma\,=\,{\rm arg}(b)$. Before writing
the explicit expressions for $U_u$ and $U_d$, it is
worth analysing the order of magnitude of the various
parameters. As we have previously shown, the parameters
$q$ and $r$ have to be small due to the mass hierarchy,
$m_u^2+m_c^2<<m_t^2$ and $m_d^2+m_s^2<<m_b^2$.
For the class of solutions we are
considering, corresponding to $\{p=0\, ;\, t=q\}$,
Eq.(2.4) for the ${\cal X}(H)$ invariant simplifies
and leads to:
$$
\sin^2({q\over 2})+\sin^2({r\over 2})+
\sin^2({q+r\over 2})=\,\,
\,\,{81\over 8}\,
{m_2^2m_3^2\over \left[m_2^2+m_3^2\right]^2} \,\, ,
\eqno(\, 3.15\, )
$$

\noindent From Eqs.(3.14) and (3.15), one obtains in leading
order:

$$
\eqalign{\eqalign{
&|q|\,\,\,{{\left[1+{r\over q}+
\left({r\over q}\right)^2\right]^{1\over 2}}}
\,\cong\,\,{9\over 2}\,{m_2\over m_3}\cr
&n\,\cong\,\sqrt{2\over 3}\,\,\,|q|\,\,\,
{{\left[1+{r\over q}+\left({r\over q}
\right)^2\right]^{1\over 2}}}\,\,\cong\,\,
3\sqrt{3\over 2}\,\,{m_2\over m_3}\cr
&\sin(\psi)\,\,\cong\,\,{\sqrt{3}\over 9}\,\, n\,\,
\cong\,\, {1\over \sqrt{2}}\,\,{m_2\over m_3}
\cr}\quad        \quad \eqalign{
&\left|{\epsilon\over n}\right|\,\,\cong\,\,
{\sqrt{3}\over 2}\,\,\left|{r\over q}\right|\,\,
{1\over {{ \left[1+{r\over q}+
\left({r\over q}\right)^2\right]^{1\over 2}}}}\cr
&\cr &\left|{a\over n}\right|\,\,\cong\,\,
{\left|1-{1\over 2}{r\over q}\right|\over
{ \left[1+{r\over q}+
\left({r\over q}\right)^2\right]^{1\over 2}}}\cr}\cr}
\eqno(\, 3.16\, )
$$
 From Eqs.(3.12), (3.13)and (3.16) it follows that in leading
order:
$$
\eqalign{
\eqalign{& |U^d_{12}|\,\,=\,\,{\sqrt{3}\over 2}\,\,
\left|{r_{_d}\over q_{_d}}\right|\cr &\cr
&|U^d_{13}|\,\,=\,\,
|U^d_{12}|\,\,|U^d_{23}|\,\,\big/2 \cr}\qquad \qquad
\eqalign{ &|U^d_{23}|\,\,=\,\,\sqrt{2}\,\,
{m_s\over m_b}\cr &\cr
&|U^d_{31}|\,\,=\,\,3\,\,|U^d_{13}|\cr}\cr}\eqno(\,
3.17\, )
$$
Analogous expressions obviously hold for $U_u$ and
therefore the CKM matrix can be readily obtained.
The formul{\ae} for $|U^d_{ij}|$, given by Eq.(3.17), show
that solutions of Class I provide a natural explanation
for the most salient features of $V_{CKM}$:

\noindent i) The almost decoupling of the third
generation is explained by the fact that
$|V_{23}|$ is proportional to $m_s/m_b$, while
$|V_{12}|$ is proportional to the ratio $r_{_d}/q_{_d}$
of two small parameters and therefore can be
considerably larger. At this stage, where we are
considering the limit $m_d=0$ ,
the ratio $r_{_d}/q_{_d}$ is an arbitrary parameter.
In the sequel, we will suggest a specific ansatz
within USY, where $m_d$ is generated and the ratio
$r_{_d}/q_{_d}$ is fixed in such a way that the
successful relation $|V_{12}|\approx (m_d/m_s)^{1/2}$.
is predicted.

\noindent ii) A hierarchy between $|V_{13}|$
and $|V_{23}|$ naturally follows since the
following relation is predicted in leading order:
$$
{|V_{13}|\over |V_{23}|}\,\,=\,\,
{1\over 2}|V_{12}| \eqno(\, 3.18\, )
$$
iii) The matrix element $|V_{31}|$ is naturally
larger than $|V_{13}|$, since in leading order
one has:
$$|V_{31}|\,\, =\,\,3\,\,|V_{13}|
\eqno(\, 3.19\, )
$$

\noindent In the above discussion we have implicitly
assumed that the dominant contribution to $V_{CKM}$
arises from $U_d$.

{\bf{4. Generating mass for the first family}}.
We have considered the limit where $m_u=m_d=0$. This
limit has the advantage of leading to a simple exact
analytical solution for $U_d$ and $U_u$, which, as we
have seen, provides an understanding of the main features
of the CKM matrix. Generating non-vanishing masses
for the first family can be viewed as a small
perturbation of $M_d$ and $M_u$, considered in Eq.(3.8). For
completeness, we give below a solution for $V_{CKM}$,
based on the parameterisation of Eq.(2.1) and
corresponding to realistic quark masses, which is in
good agreement with our experimental knowledge on
$V_{CKM}$.
In this solution the value of the parameters are
near to those of Eq.(3.8), corresponding to the
the limit $m_u=m_d=0$.
We present below $|V_{CKM}|$ and also the value of
the invariant
$J={\rm Im}(V_{12} V_{23} V_{22}^* V_{13}^*)$,
which measures the strength of CP violation
in the SM. In the evaluation of $V_{CKM}$, we have
considered
non-vanishing phases in the diagonal unitary matrices
$K_u$ and $K_d$, defined in Eq.(2.2). It turns
out that this is crucial in order to obtain a sufficiently
large value of $|J|$, assuming that $\epsilon$ only
receives contributions from the Kobayashi-Maskawa (KM)
mechanism.

\noindent Input:
$$
\matrix{p_{_d}=0.&q_{_d}=0.1140&r_{_d}=0.0498&
t_{_d}=0.1482\cr p_{_u}=0.&
q_{_u}=0.0128&r_{_u}=0.0112&t_{_u}=0.0130\cr}\quad ;\quad
\psi_1=0.020\,\, ,\,\,\psi_2 =0.058\eqno(\, 4.1\,)
$$
Output:
$$\matrix {m_u(1\, GeV)&=5.3 \, MeV &
m_c(1\,GeV)=1.35\, GeV &
m_t(1\, GeV)=290 \, GeV \cr
m_d(1\, GeV)&=9.0 \, MeV & m_s(1\, GeV)=190\, MeV
&
m_b(1\, GeV)= 5.2\, GeV \cr} \eqno(\, 4.2\, )
$$

$$
|V_{CKM}| \, =\,\,\left[\matrix{0.9754 & 0.2203
&0.0028\cr 0.2201 & 0.9748 &0.0350 \cr
0.0103 & 0.0336 & 0.9994 \cr}\right]\quad ;\quad
\left|{V_{13}\over V_{23}}\right|\,=\, 0.081\quad ;
\quad |J|\,=\, 0.8\cdot 10^{-5}
\eqno(\, 4.3  \, )
$$

{\bf{5. Predicting ${\bf V_{CKM}}$ in
terms of quark mass ratios}}. In this section, we will
propose an ansatz in the USY framework, where $V_{CKM}$
is determined in terms of quark mass ratios. In looking
for an ansatz, we will be guided by the results we
obtained
in the limiting case where the first generation is
massless. We will assume that the down quark acquires
mass through a small perturbation of the matrix $M_d$
considered in Eq.(3.8). and we propose:

$$
M_d\,=\,\, c_d\,\,
\left(\matrix{1&e^{ir_{_d}}&1\cr & & \cr
e^{iq_{_d}}&1&e^{i(q_{_d}+r_{_d})}\cr & & \cr
1&1&1\cr}\right)\eqno(\, 5.1 \, )
$$
 From Eqs.(2.4) and (5.1), we get the following exact
relation:
$$
\left[\sin({r_{_d}\over 2})\right]^4\,\,=
\,\,{3^6\over 2^4}\,\,
{m_d^2\, m_s^2\, m_b^2\over
[m_d^2+m_s^2+m_b^2]^3}\eqno(\, 5.2 \, )
$$
In this ansatz, the value of $|r_{_d}|$ is exactly
determined by the quark mass ratios through Eq.(5.2).
Given the mass hierarchy, the following approximate
relation holds:
$$
|r_{_d}|\,\,\cong\,\, 3\sqrt 3\,\,
{\sqrt{m_d\, m_s}\over m_b} \eqno(\, 5.3 \, )
$$
Similarly, the value of $|q_{_d}|$ can be obtained
from the quark mass ratios, by using the exact
relation:
$$
\left[\sin({q_{_d}+r_{_d}\over 2})
\right]^2\,\,=\,\,{D_{ _{\cal X}}-
3D_{ _{\delta}}\over 2(1-D_{ _{\delta}} )}\eqno(\, 5.4\,
)
$$
with:
$$
D_{ _{\cal X}} ={3^4\over 2^3}\,{m_d^2m_s^2+
m_d^2m_b^2+m_s^2m_b^2\over [m_d^2+m_s^2+m_b^2]^2}
\quad ;\quad D_{ _{\delta}}={3^3\over 2^2}\,
{m_d\, m_s\, m_b\over
[m_d^2+m_s^2+m_b^2]^{3\over 2}}\eqno(\, 5.5\, )
$$
 From Eqs.(5.3) and (5.4) it follows that:
$$
\left|{ r_{_d}\over q_{_d}}\right|\,\,\cong\,\,
{2\over\sqrt 3}\,\,\sqrt{m_d\over m_s}\eqno(\, 5.6\, )
$$
Recall that the ratio $|r_{_d}/ q_{_d}|$ was an
arbitrary parameter in the analysis of the limiting
case $m_u=m_d=0$, of Eq.(3.8). On the contrary, in the
specific ansatz of Eq.(5.1), $|r_{_d}/ q_{_d}|$ is fixed
by a ratio of quark masses.
The fact that $q_{_d}$ and $r_{_d}$ can be expressed in
terms of quark mass ratios, enables one to easily
diagonalize the quark mass matrices. As before, we make
first the weak-basis transformation of Eq.(3.9), with
$F$ given by Eq.(3.10). Consider now the eigenvalue
equation:
$$
\left({\tilde H'}_d-\lambda_i\,\, 1\right)\,\,
{\overrightarrow u}_i\,\,=\,\, 0\eqno(\, 5.7\, )
$$
where ${\tilde H'}_d$ are the dimensionless
Hermitian matrices in the
new basis ${\tilde H'}_d\,\,=\,\, F^\dagger\,\,
{\tilde H}_d\,\, F$, with
${\tilde H}_d=(1/3c_d^2)(M_d\,\,M_d^\dagger)$.
Since ${\tilde H'}_d$ in the
present ansatz can be expressed in terms of quark mass
ratios by using Eqs.(5.2), (5.4) and (5.5)
an exact solution of the
eigenvalue equation can be easily found. Indeed from
Eq.(5.7) one has:
$$
{\overrightarrow u}_i \,\,=\,\,{1\over N_i}\,\,
\left({\overrightarrow x}_i \,\times\,
{\overrightarrow y}_i\right)\eqno(\, 5.8\, )
$$
where the $N_i$ are normalisation factors and:
$$
\left({\overrightarrow x}_i \, ,\,
{\overrightarrow y}_i\, ,\,{\overrightarrow z}_i
\right)\,\,=\,\,
\left({\tilde H'}_d-\lambda_i\,\, 1\right)^T\eqno(\,
5.9\, )
$$
 From the Eqs.(5.1),(5.2),(5.4),(5.5) and (5.7)-(5.9)
one gets an exact solution for the
unitary matrix $U_d$, which diagonalizes the down
quark mass matrix. Expanding $U_d$ in terms of the
quark mass ratios one obtains in leading order:
$$
\eqalign{
\eqalign{& |U^d_{12}|\,\,=\,\,\sqrt{m_d\over m_s}
\cr &\cr&|U^d_{13}|\,\,=\,\,{1\over\sqrt 2}\,\,
\sqrt{m_d\, m_s\over m_b^2}  \cr} \qquad\qquad
\eqalign{ &|U^d_{23}|\,\,=\,\,\sqrt{2}\,\,
{m_s\over m_b}\cr &\cr
&|U^d_{31}|\,\,=\,\,{3\over\sqrt 2}\,\,
\sqrt{m_d\, m_s\over m_b^2}\cr}\cr}\eqno(\, 5.10\, )
$$
Comparing these results with those of Eq.(3.17), it is seen
that the values of $U^d_{ij}$ given by Eq.(5.10) can be
obtained from Eq.(3.17), by simply putting the ratio
$| r_{_d}/ q_{_d}|\,=\,
(2/\sqrt 3)\,\,(m_d/ m_s)^{1/2} $,
as predicted in the present ansatz in Eq.(5.6).
This was, in a certain sense, to be expected,
since $M_d$ in Eq.(5.1) can
be viewed as the result of a small perturbation of
$M_d$ in Eq.(3.8), whose main effect is generating a down
quark mass and fixing the ratio $r/q$. However, to
leading order, the structure of $U_d$ is not changed.
It is remarkable that the factor
$2/\sqrt 3$ in Eq.(5.6) just cancels the factor
$\sqrt 3/ 2$ in the expression for $U_d$ in Eq.(3.17),
so that one obtains the correct prediction
$|U^d_{12}|\,=\, (m_d/m_s)^{1/2}$.
In order to derive the predictions for $V_{CKM}$,
we need to specify the structure of $M_u$. If one
takes for $M_u$ the same ansatz we have chosen for $M_d$
in Eq.(5.1), one obtains
a good fit for all elements of $V_{CKM}$, with the
possible exception of $V_{12}$.
The potential difficulty with $V_{12}$ has to do
with the fact that, for the above choice for $M_u$, one
obtains in leading order
$|V_{12}|=(m_d/ m_s)^{1/2}\,\,
\pm\,\,(m_u/ m_c)^{1/2}$, where the sign
ambiguity results from the ambiguity in the relative
sign of $q_{ _d}$, $r_{ _d}$, $q_{ _u}$ and $r_{ _u}$,
as extracted from Eqs.(5.2), (5.4) and its equivalent for
the up
quarks. From the experimental limit on $(m_u/m_d)$,
$(m_d/m_s)$, $m_s$ and $m_c$, it can be verified that
the experimental value $|V_{12}|=0.2205\,\,\pm\,\,
0.0018$ [6] can be accomodated only if one takes for $m_u$
a value smaller than what is favoured by most of
the analyses [6] or if one chooses a different ansatz for
$M_u$. Therefore, this difficulty can be avoided
in two ways:

\noindent i) One may choose a ratio $m_u/m_d$ smaller
than the standard analysis [7]. Indeed there is some
controversy [8] about the actual value of $m_u$. Taking
$m_u=0$ would have the attractiveness of providing a
simple solution to the strong CP problem [9]. If we
choose $m_u=1.0\,\, MeV\,\,(1\,\, GeV)$ and take as
ans"tze for both $M_d$ and $M_u$ the form of Eq.(5.1),
we get the following prediction for $|V_{CKM}|$:

$$
|V_{CKM}| \, =\,\,\left[\matrix{0.9752 & 0.2212
&0.0029\cr 0.2210 & 0.9745 &0.0385 \cr
0.0113 & 0.0369 & 0.9993 \cr}\right]\qquad ;\qquad
\left|{V_{13}\over V_{23}}\right|\,=\, 0.075
\eqno(\, 5.11\,)
$$
where the other quark masses were chosen to be within
the experimentally allowed  range:
$$\matrix {m_u(1\, GeV)&=1.0 \, MeV &
m_c(1\,GeV)=1.35\, GeV &
m_t(1\, GeV)=300 \, GeV \cr
m_d(1\, GeV)&=6.3 \, MeV & m_s(1\, GeV)=160\, MeV
&
m_b(1\, GeV)= 5.6\, GeV \cr} \eqno(\, 5.12\,)
$$

\noindent ii) There is no fundamental reason for the
choosing the same ansatz for $M_u$ and $M_d$. In fact,
various of the viable Yukawa textures recently classified in
Ref.[10] correspond to taking different forms for
$M_u$ and $M_d$. Encouraged by the results we have
obtained for $|U_d|$ in Eq.(5.10), we propose the
following ansatz:

$$
M_d\,=\,\, c_d\,\,
\left(\matrix{1&e^{ir_{_d}}&1\cr & & \cr
e^{iq_{_d}}&1&e^{i(q_{_d}+r_{_d})}\cr & & \cr
1&1&1\cr}\right)\quad ;\quad
M_u\,=\,\, c_u\,\,
\left(\matrix{e^{ip_{_u}}&1&1\cr & & \cr
e^{iq_{_u}}&1&e^{iq_{_u}}\cr
& & \cr1&1&1\cr}\right)\eqno(\, 5.13\,)
$$
We have thus kept the same $M_d$ as in Eq.(5.1),
but have taken a different form for $M_u$. Note that
in this ansatz both $M_u$ and $M_d$ have the same
form in a specific limit where $m_u=m_d=0$, which
corresponds to having $p_{ _u}=r_{ _d}=0$, but
both $q_{ _u}$ and  $q_{ _d}$ non-vanishing. It is
only when a mass is generated for the first family that
an asymmetry arises between $M_u$ and $M_d$.
It can be verified that $U_u$ for the above ansatz
can also be expressed in terms of quark mass ratios.
As a result $V_{CKM}$ has no free parameters. The
ansatz predicts for the CKM matrix:
$$
|V_{CKM}| \, =\,\,\left[\matrix{0.9753 & 0.2207
&0.0036\cr 0.2203 & 0.9744 &0.0443 \cr
0.0133 & 0.0424 & 0.9990 \cr}\right] \qquad ;\qquad
\left|{V_{13}\over V_{23}}\right|\,=\, 0.082
\eqno(\, 5.14\,)
$$
where we have taken the following quark masses:
$$\matrix {m_u(1\, GeV)&=4.0 \, MeV &
m_c(1\,GeV)=1.35\, GeV &
m_t(1\, GeV)=290 \, GeV \cr
m_d(1\, GeV)&=6.6 \, MeV & m_s(1\, GeV)=133\, MeV
&m_b(1\, GeV)= 5.7\, GeV \cr} \eqno(\, 5.15\,)
$$
The values predicted for $|V^{CKM}_{ij}|$ are in good
agreement with experiment. The crucial difference
between this ansatz and the one with both $M_d$ and $M_u$
of the form of Eq.(5.1) is that now
$|U^u_{12}|\,=\,(1/\sqrt{3})\,(m_u/m_c)$ and therefore
one obtains:
$$
|V^{CKM}_{12}|\,\,=\,\,\left( {m_d\over m_s}\right)^{1\over
2}-
{1\over 2}\left( {m_d\over m_s}\right)^{3\over 2}+
{1\over \sqrt 3}\left({m_u\over m_c}\right)\eqno(\, 5.16\, )
$$
where we have kept the subleading contribution arising
from $U_d$, since its size is comparable to the leading
contribution from $U_u$.
This new formula for $V^{CKM}_{12}$
is the essential reason why the ansatz of Eq.(5.13)
leads to all $|V^{CKM}_{ij}|$ in agreement with experiment,
for values of quark masses within the allowed ranges.

At this stage, the following comment is in order.
In the USY framework the
strength of CP violation, as measured by the rephasing-
invariant $J$, can in general be in agreement with the
experimental value of $\epsilon$, as it was illustrated
by the example of Eq.(4.3). However, in the specific ansatz
which we have proposed, the strength of CP violation
through the KM mechanism is not sufficient to account
alone for the observed CP violation in the Kaon sector.
This should not be considered
a drawback of the present ansatz. In most
extensions of the SM there are new contributions to
$\epsilon$, the simplest example occurring in
models with more than one Higgs doublet [11]. In fact,
new sources of CP violation beyond the SM are needed
[12] in order to generate the observed baryon asymmetry
at the electroweak phase transition.

 {\bf{ 6. Summary an Conclusions}}. We have studied the
USY hypothesis in the limit where $m_u=m_d=0$. It was
shown that all solutions with a vanishing mass for the
first generation fall into two classes.
In one of these classes the main features of the
CKM matrix are correctly predicted.
Inspired by this analysis, we proposed, within the USY
framework, a specific ansatz with a high predictive
power,
where the CKM elements are determined in terms of
quark mass ratios, with no free parameters. The
predictions for all moduli of the CKM matrix elements are in
agreement with experiment.

In conclusion, we find it remarkable that a simple physical
idea such as the USY hypothesis can lead to a highly
successful ansatz. This provides motivation to address
the deeper question of finding a symmetry principle
which can lead to the universality of Yukawa couplings.

\bigskip
{\bf Acknowledgments}

We would like to thank Lincoln Wolfenstein
for interesting discussions. This work was supported by Science
Project No. SCI-CT91-0729 and EC contract No. CHRX-CT93-0132.

\vfill\eject
{\bigtenrm{ References}}
\bigskip
\item {[1]} G. C. Branco, J. I. Silva-Marcos and M. N. Rebelo,
Phys. Lett. B 237 (1990) 446.
\item {[2]} S. L. Glashow and S. Weinberg, Phys. Rev. D 15 (1977) 1958.
\item {[3]} P. M. Fishbane and P. Kaus, Phys. Rev. D 49 (1994) 4780;
\item {}J. Kalinowski and M. Olechowski, Phys. Lett. B 251 (1990) 584.
\item {[4]} H. Harari, H. Haut and J. Weyers, Phys. Lett. B 78 (1978) 459;
\item {} Y. Chikashige, G. Gelmini, R. D. Peccei and M. Roncadelli,
\item{}Phys. Lett. B 94 (1980) 499;
\item {} C. Jarlskog in Proc. of the Int. Symp. on Production and
Decay of Heavy Flavours, Heidelberg, Germany, 1986;
\item {}P. Kaus and S. Meshkov, Mod. Phys. Lett. A 3 (1988) 1251;
\item {}{\it ibid}, Mod. Phys. Lett. A 4 (1989) 603;
\item {}G. C. Branco, J. I. Silva-Marcos and M. N. Rebelo, Ref.[1] ;
\item {}H. Fritzsch and J. Plankl, Phys. Lett. B 237 (1990) 446.
\item {[5]} H. Fritzsch and J. Plankl in Ref.[4];
\item {}H. Fritzsch and Z. Xing, Max-Plack preprint MPI-PhT/95-08 (1995).
\item {[6]} Review of Particle Properties, Phys. Rev. D 50 (1994).
\item {[7]} J. Gasser and H. Leutwyler, Phys. Rep. 87 (1982) 77.
\item {[8]} K. Choi, Nucl. Phys. B 383 (1992) 58.
\item {[9]} For excellent reviews, see: R. D. Peccei,
DESY Report 88-109 (1988)in
\item{}"CP violation", World Scientific, Singapore;
\item {}J. Kim, Phys. Rep. 150 (1987);
\item {}H. Y. Cheng, Phys. Rep. 158 (1988).
\item {[10]} P. Ramond, R. G. Roberts and G. G. Ross, Nucl. Phys. B 406 (1993)
19.
\item {[11]} G. C. Branco and M. N. Rebelo, Phys. Lett. B 160 (1985) 117;
\item {}J. Liu and L. Wolfenstein, Nucl. Phys. B 289 (1987) 1.
\item {[12]} M. B. Gavela, P. Hernandez, J. Orloff and O. PŠne, Mod. Phys.
Lett. A 9 (1994);
\item {}{\it ibid}, Nucl. Phys. B 430 (1994) 382;
\item {}P. Huet and E. Sather, Phys. Rev. D 51 (1995) 379.

\bye